\documentclass[times,twocolumn,final]{elsarticle}

\usepackage{medima}
\usepackage{framed,multirow}

\usepackage{amssymb}
\usepackage{latexsym}

\usepackage{url}
\usepackage{xcolor}

\usepackage{hyperref}
\usepackage{subcaption}
\usepackage{booktabs}
\usepackage{colortbl}
\usepackage{amsmath}
\definecolor{newcolor}{rgb}{.8,.349,.1}
\definecolor{mygray}{gray}{0.9}

\journal{Medical Image Analysis}

\begin{document}

\verso{Yunhe Gao \textit{et~al.}}

\begin{frontmatter}

\title{FocusNetv2: Imbalanced Large and Small Organ Segmentation with Adversarial Shape Constraint for Head and Neck CT Images}

\author[1,2,4]{Yunhe Gao}
\author[3]{Rui Huang}
\author[1]{Yiwei Yang}
\author[1]{Jie Zhang}
\author[1]{Kainan Shao}
\author[1]{Changjuan Tao}
\author[1]{Yuanyuan Chen}
\author[2]{Dimitris N. Metaxas}
\author[4]{Hongsheng Li\corref{cor1}}
\author[1]{Ming Chen\corref{cor1}}
\cortext[cor1]{Corresponding author\\ \indent \indent \textit{e-mail:} \texttt{chenming@zjcc.org.cn} (Ming Chen), \\ 
\indent \indent \indent \indent \indent \, \texttt{hsli@cuhk.edu.hk} (Hongsheng Li)}

\address[1] {Cancer Hospital of University of the Chinese Academy of Sciences (Zhejiang Cancer Hospital), China}
\address[2]{Department of Computer Science, Rutgers University, Piscataway, New Jersey, USA}
\address[3]{SenseTime Research, China}
\address[4]{Department of Electronic Engineering, The Chinese University of Hong Kong, Hong Kong SAR, China}

\received{24 Jan. 2020}
\finalform{}
\accepted{}
\availableonline{}
\communicated{}

\begin{abstract}
Radiotherapy is a treatment where radiation is used to eliminate cancer cells. The delineation of organs-at-risk (OARs) is a vital step in radiotherapy treatment planning to avoid damage to healthy organs. For nasopharyngeal cancer, more than 20 OARs are needed to be precisely segmented in advance. The challenge of this task lies in complex anatomical structure, low-contrast organ contours, and the extremely imbalanced size between large and small organs. Common segmentation methods that treat them equally would generally lead to inaccurate small-organ labeling. We propose a novel two-stage deep neural network, FocusNetv2, to solve this challenging problem by automatically locating, ROI-pooling, and segmenting small organs with specifically designed small-organ localization and segmentation sub-networks while maintaining the accuracy of large organ segmentation. In addition to our original FocusNet, we employ a novel adversarial shape constraint on small organs to ensure the consistency between estimated small-organ shapes and organ shape prior knowledge. Our proposed framework is extensively tested on both self-collected dataset of 1,164 CT scans and the \emph{MICCAI Head and Neck Auto Segmentation Challenge 2015} dataset, which shows superior performance compared with state-of-the-art head and neck OAR segmentation methods.
\end{abstract}

\end{frontmatter}


\section{Introduction}

Radiation therapy is an important treatment for cancers. High energy radiation beams are focused on the tumor area to prevent tumor cell division, and ultimately result in tumor cell death. However, radiation is not specific to cancer cells and can also damage healthy cells. For nasopharyngeal carcinoma, more than 20 organs-at-risk (OARs) in the head and neck (HaN) region may be affected during radiotherapy, which may cause side effects including dysphagia, xerostomia, hypopsia, dysacusis radiation-induced lower cranial neuropathy and etc. Therefore, during radiotherapy treatment planning, radiologists need to accurately plan the radiotherapy path to ensure that the radiation dose received by normal organs is within safe limits.

The quality of organs-at-risk delineation is a core factor affecting the efficacy and side effects of radiotherapy. Clinically, radiologists have to spend several hours manually delineating organs. It is usually very time consuming and requires a high level of professionalism for the radiologist. In some underdeveloped areas, qualified radiologists are very scarce resources. Thus designing a high-performance and robust OAR segmentation algorithm can effectively alleviate this dilemma, reduce the workload of doctors, improve the quality of radiotherapy, and reduce the waiting time for patients, which would greatly benefit both patients and doctors.

\begin{figure}[tb!]
\centering
    \includegraphics[height=6.5cm]{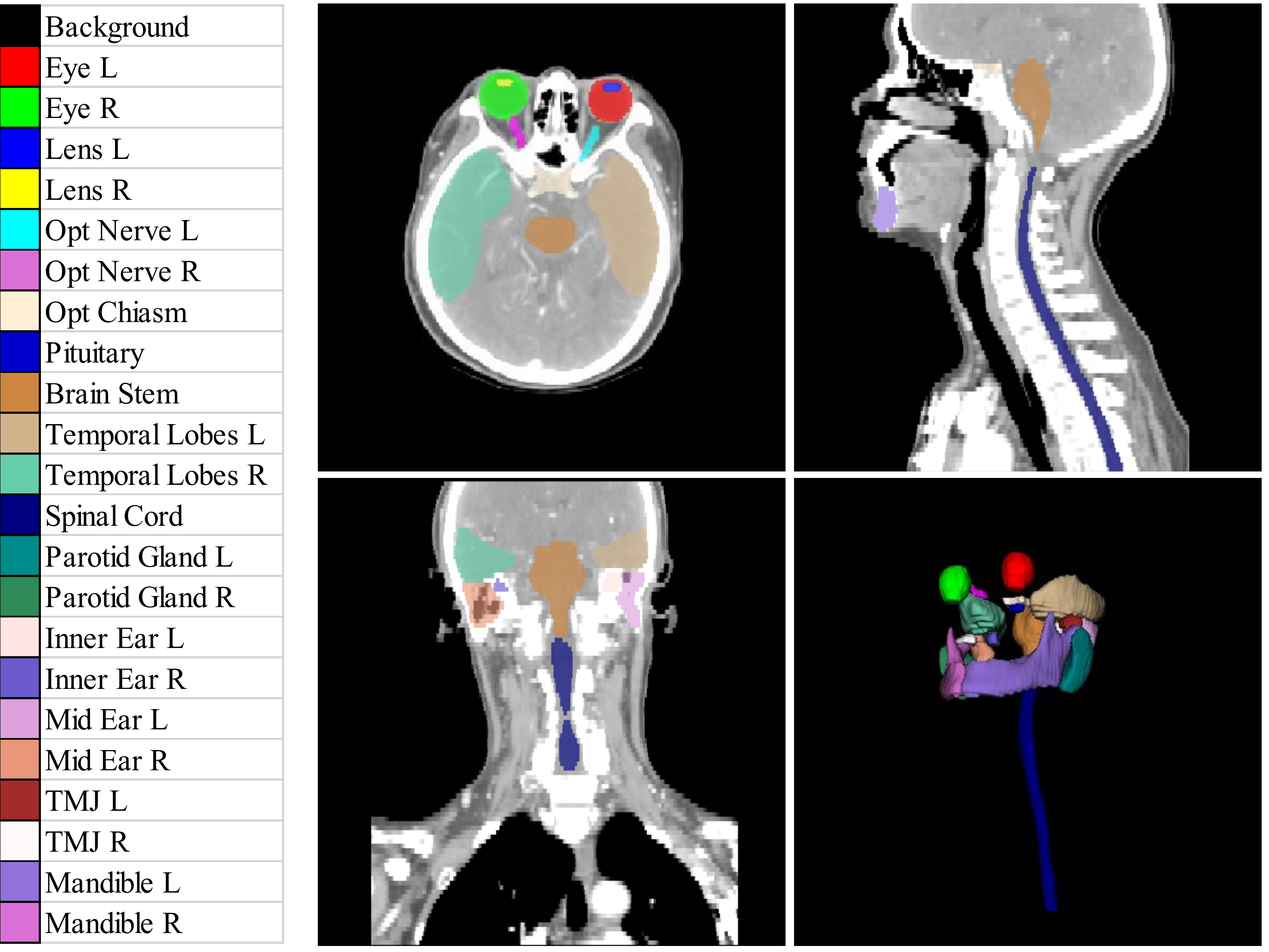} \\
    \caption{Over 20 OARs are an in radiotherapy of nasopharyngeal cancer. Visualization of 22 OARs' annotations in our large-scale self-collected dataset. The key difficulty of this task lies in complex anatomical structure, low contrast of soft tissue, and extreme unbalanced samples among large and small organs. }\label{fig:3d_fig}
    
\end{figure}

The main difficulty of this task lies in the following aspects. First, the complex anatomical structure, over 20 OARs in HaN have various structures and complex shapes. For example, the optic chiasm is not a smooth convex shape, but an X-like shape. Second, due to the limitation of CT imaging, the contrast of soft tissue is relatively low. It usually cannot clearly show the boundaries of organs and makes the automatic localization of organs' boundaries a challenging task. Moreover, the sizes of organs in the head and neck are extremely unbalanced. The ratio of sizes between large and small organs can reach hundreds, e.g., the parotid gland occupies tens of thousands of voxels, while pituitary gland only occupies about one hundred voxels\footnote{Small organs are defined as organs that have small sizes. In our collected dataset, small organs include left and right lens, left and right optic nerve, optic chiasm and pituitary.}.

Over the past decade, there have been many approaches proposed to resolve the challenging problem of HaN OAR segmentation. Atlas-based methods were commonly used where there are only a small number of annotated images available. However, they are based on image registration techniques and might generate incorrect organ delineations, especially when tumors occupy the organs. The atlas-based methods' time cost can take up to tens of minutes as it is computationally intensive. Recently, convolutional neural networks (CNN), with its powerful feature representation capability, have made revolutionary progress in OAR segmentation \citep{ibragimov2017segmentation, raudaschl2017evaluation, fritscher2016deep, wang2017hierarchical, ren2018interleaved, zhu2019anatomynet}. However, existing segmentation CNNs are not optimized for unbalanced organ segmentation tasks. These networks generally produce accurate segmentation maps for large organs, while the accuracy of small organs is often sacrificed. Furthermore, trained with per-pixel loss, such as cross-entropy loss, existing deep learning methods can not guarantee consistency between predict organ shapes and prior knowledge.

In order to solve the problems mentioned above, we observe how professional doctors delineate OARs. For large organs, they usually label them at regular scales. For small organs, they first localize the rough location and zoom in for accurate delineation. For organs with blurry boundaries in CT images, the doctors usually fit a shape on the image based on prior medical knowledge. Take the optic chiasm for an example, doctors usually fit an X-shaped label to the image as much as possible at its rough location. According to this observation, we propose a novel 3D convolutional neural network, FocusNetv2, which is delicately designed for accurate segmentation of both large organs and small organs in HaN CT scans. To better regularize the predicted organ shapes, in addition to our original FocusNet \citep{gao2019focusnet}, we further propose a novel adversarial autoencoder (AAE) to incorporate shape constraint as extra supervisions for the segmentation network.

The overall framework of our approach is shown in Fig. \ref{fig:overview}. In particular, our network has two main components: the segmentation network and the adversarial autoencoder (AAE) for organ shape constraint. The segmentation network solves the extremely unbalanced data in a two-stage framework, which consists of three parts: main segmentation network (S-Net), Small-Organ Localization branch (SOL-Net) and Small-Organ Segmentation branch (SOS-Net). It imitates the process of how doctors delineate medical images. The segmentation network first segments all organs with the main segmentation network (S-Net) and localizes the central locations of a series of pre-defined small organs with the Small-Organ Localization branch (SOL-Net). Multi-scale features and high-resolution images are ROI-pooled by the Small-Organ segmentation branch (SOS-Net) to generate small-organ label maps. After further adding shape constraints with the proposed adversarial autoencoder (AAE), it encourages the predictions of the segmentation network being consistent with the prior shapes of different organs, even if there are no clear boundaries in the CT images. To the best of our knowledge, this is the first segmentation method that leverages both the autoencoder and adversarial learning for shape regularization.

This paper is an extension of our preliminary work, FocusNet \citep{gao2019focusnet} (denoted as FocusNetv1 in the following paper). Several modifications were made to both methodology and experiments, including modeling organ shape priors by the newly proposed adversarial autoencoder, additional experiments on larger datasets, and more ablation studies. The rest of this paper is organized as follows. Section 2 reviews related works on semantic segmentation in medical images, OAR segmentation and shape constraints by CNNs. Section 3 describes the proposed OAR segmentation framework and the adversarial shape autoencoder for better organ-shape regularization. Section 4 presents the experimental results. At last, Section 5 concludes the methodology and experiments.

\section{Related Work}

\subsection{CNNs for medical image segmentation}

Recently, convolutional neural networks have greatly advanced the field of medical image analysis because of its ability to learn more representative features from data. CNNs demonstrate state-of-the-art performance in many challenging tasks, such as image classification, segmentation, detection, registration, super-resolution and etc.


\citet{long2015fully} first proposed fully convolutional network (FCN), which uses convolutions with 1$\times$1 sized filter to replace the fully connected layer, and allows the prediction of multiple pixels at the same time. \citet{ronneberger2015u} further built a ``U'' shaped network (named U-Net) with a contracting path and a symmetric expanding path. Skip connections are also used to propagate features from early layers to later ones. A vast number of works based on variants of FCN and U-Net are applied in the field of 2D medical image segmentation \citep{christ2016automatic, yi2019attentive, brosch2016deep, roth2016spatial, tan2018deep}.

For 3D images like CT or MR, 2D CNNs can be utilized in a slice-by-slice manner, however, the context information encoded in the volumetric data is ignored. Some 2.5D methods \citep{roth2014new, xu2017neonatal} attempted to incorporate 3D spatial information by using three orthogonal slices or adjacent slices. But their representation capabilities are still limited by 2D convolutional kernels. To overcome this weakness, 3D CNN-based algorithms are proposed. For example, the 3D version of U-Net is proposed by \citet{cciccek20163d}; \citet{milletari2016v} proposed V-Net, which introduces the residual connection \citep{he2016deep} between building blocks to mitigate gradient vanishing problem. Several 3D networks were also proposed for different applications, such as \citet{merkow2016dense, dou20163d, kamnitsas2017efficient}. 

Even though 3D CNN based methods can better leverage spatial context for learning better feature representations, the sample imbalanced problem is magnified in 3D tasks, as the training errors are mostly dominated by voxels belong to large organs. \citet{ronneberger2015u} proposed to use the weighted cross-entropy loss function, while \citet{milletari2016v} proposed the Dice coefficient loss, they can only mitigate the challenge of unbalanced data but are far away from solving it.

\begin{figure*}[hbt!]
\centering
    \includegraphics[height=7cm]{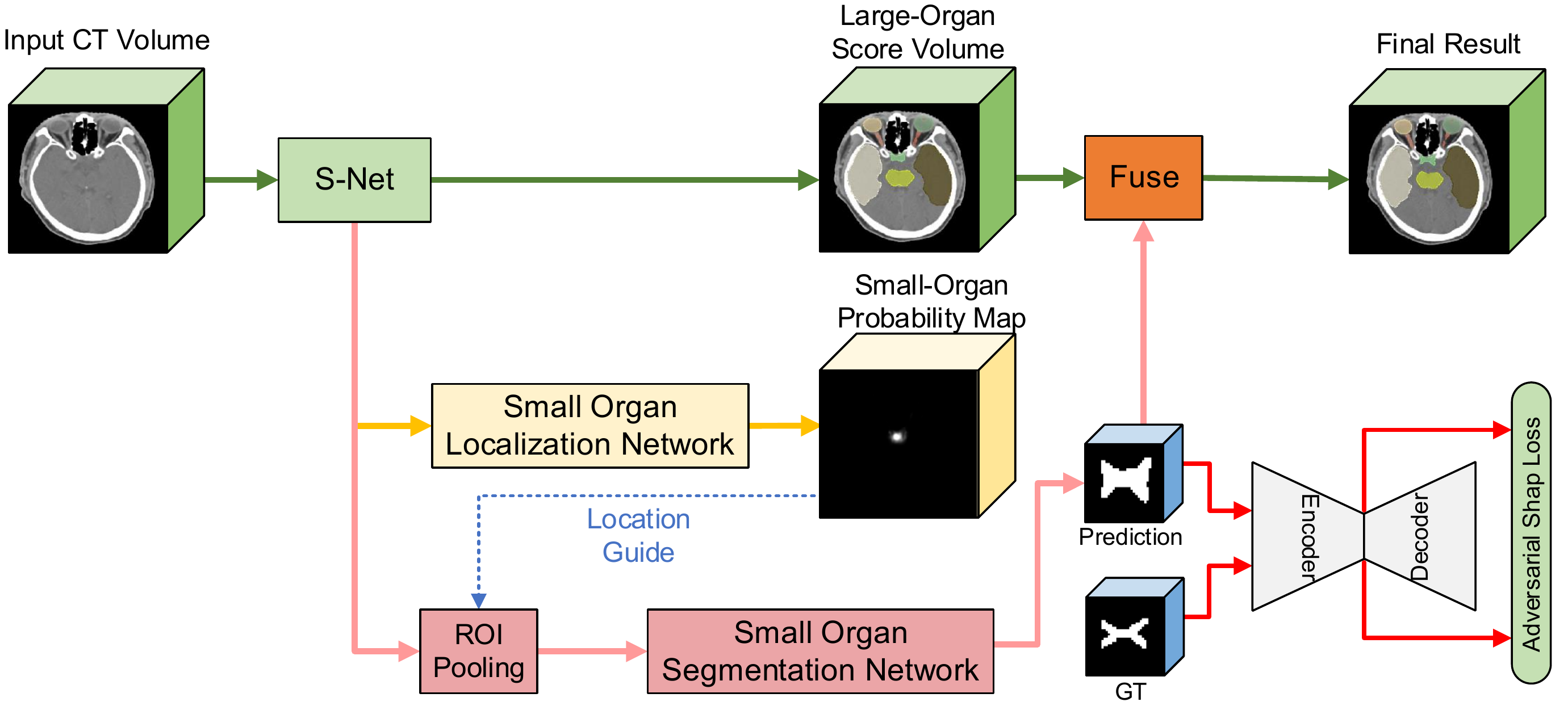} \\
    \caption{Overall framework of out proposed FocusNetv2.}\label{fig:overview}
    
\end{figure*}

\subsection{OAR segmentation for head and neck region}

There are numerous works of OAR segmentation proposed for radiotherapy treatment planning of different body parts. Atlas-based methods are among the most commonly used traditional approaches. The optimal transformation between the atlas, which has the pre-segmented annotation map, and the image to be segmented is aligned by affine and deformable registration. Then the segmentation for the target image can be obtained by applying this transformation on the annotation map of the reference image. The reference images can be multiple ones with expert annotations or templates generated from the training set. The accuracy of atlas-based approaches is affected by two factors: First, the capability of the registration method, whether it can align the target image and atlas images accurately. Different approaches, such as Demons registration \citep{thirion1998image, qazi2011auto}, block-matching \citep{ourselin2000block, han2008atlas}, and B-Spline registration \citep{zhang2009segmentation}, have been proposed. Second, physiologic or pathologic anatomical variations of some organs make it difficult to find the optimal correspondences between target images and reference images, thus some methods are proposed to use atlases that reflect average patient anatomy \citep{commowick2009using} or the fusion of results from multi-atlas \citep{commowick2007efficient, rohlfing2004evaluation}. Some hybrid methods post-process results from atlas-based segmentation using active contours \citep{zhang2009segmentation} and graph cuts \citep{van2011automated, fortunati2013tissue}. Although atlas-based methods have the advantages of robustness and can perform segmentation without user interaction, they are based on image registration techniques and might generate incorrect organ maps if the organs are occupied by tumors. The time cost can be up to tens of minutes due to its huge amount of computation.

Recently, convolutional neural networks were adopted to advance OAR delineation accuracy significantly . \citet{ibragimov2017segmentation} proposed the first deep learning-based algorithm. They first detect OARs by the head center point and then train a patch-based CNN to classify voxels in the region of interest. \citet{ren2018interleaved} proposed an interleaved 3D-CNN for the joint segmentation of small organs in HaN, where the region of interest is obtained via registration techniques. \citet{zhu2019anatomynet} proposed a 3D squeeze-and-excitation U-Net for fast segmentation. \citet{tong2018fully}  proposed a fully convolutional neural network with a shape representation model. \citet{tang2019clinically} also proposed a two-stage segmentation method based on detection, where local contrast normalization is applied in the segmentation head for achieving better segmentation performance. There are also methods \citep{mlynarski2020anatomically} segment OARs in other modalities, such as MRI.

\subsection{Shape regularization in segmentation CNNs}

Current CNN-based approaches are usually trained with pixel-wise losses, such as the cross-entropy loss and the dice loss, which consider pixels' prediction errors separately. Even with large enough receptive fields, they cannot maintain the overall shapes or higher-level structures of the organ of interest. Conditional random field (CRF) and graph cut methods were proposed to enforce spatial contiguity of the output label maps. Several approaches attempted to incorporate shape constraints in CNNs, \citet{mosinska2018beyond} incorporated a pre-trained VGG network to extract higher-order topological features from the network's predictions and ground truth label maps, and then minimizes the $L2$ loss between them. However, the ImageNet pre-trained VGG network may not capture various organs' anatomy in medical images. \citet{oktay2017anatomically} used an autoencoder (AE) to learn representations of shapes from ground-truth annotations and minimize the Euclidean loss between the encoded latent codes of the network's predicted label maps and ground truth label maps. \citet{tong2018fully} applied similar ideas with \citet{oktay2017anatomically} in the segmentation of OARs. \citet{al2018spnet} modified the UNet to generate a signed distance function (SDF) instead of segmentation maps and computes the errors directly in the shape domain using principal component analysis (PCA). 
Adversarial losses were also introduced as high-order shape regularization \citep{yang2017automatic, xue2018segan}. 
The basic idea of methods mentioned above are similar, i.e., instead of measuring the shape similarity with ground-truth in the pixel domain, they proposed to measure the similarity in a lower dimension manifold by different kinds of projection (i.e., pre-trained VGG network, autoencoder, PCA or discriminator). Therefore, the key is the quality of the shape projection. However, existing adversarial discriminators lack generalization capability. The autoencoder in \citet{oktay2017anatomically,tong2018fully} was trained with only ground-truth annotation. It cannot well encode various predicted shapes by the segmentation models. To mitigate the problem, we proposed a novel adversarial shape constraint based on autoencoder that incorporates benefits from both autoencoder and adversarial discriminators.

There are three main differences between this work and the previous work. First, we designed a high-performance backbone network with fewer down-samplings but using DenseASPP \citep{Yang_2018_CVPR} to retain more detailed high-resolution information as well as learn multi-scale features. Second, we use a two-stage segmentation model for small organs to solve the imbalance problem between large and small organs. The third and most crucial difference is that we use the adversarial autoencoder to apply shape constraints to the network prediction, which allows our network to generate predicted shapes that conform to prior medical knowledge for specific organs with unclear boundaries and low contrast. In practical clinical applications, our model can generate delineation results that are more acceptable by human doctors to improve the efficiency of radiotherapy treatment planning.

\section{Method}

The overall structure of the proposed FocusNetv2 is illustrated in Fig. \ref{fig:overview}. The FocusNetv2 contains two main parts, the segmentation network and the adversarial autoencoder for shape regularization. They are trained in an adversarial fashion. For the segmentation network, it imitates how human doctors delineate medical images: labeling large organs at a regular scale, while for the small organs, they first localize them and then zoom in for further accurate delineation. Therefore, the segmentation network first segments all organs with the main segmentation network (S-Net) and localizes the small-organ central locations with the Small-Organ Localization branch (SOL-Net). Multi-scale features and raw CT images are ROI-pooled from small-organ locations by the Small-Organ Segmentation branch (SOS-Net) to generate small-organ label maps. Therefore, the proposed segmentation network can solve the class imbalance for HaN OAR segmentation.

To better regularize the shapes output by the segmentation network, we proposed an adversarial autoencoder (AAE) to constrain the estimated small-organ shapes. The AAE is trained with two types of inputs, ground truth label map and predicted mask from our SOS-Net. The segmentation network and the AAE are alternately trained in an adversarial fashion. The AAE tries to encode better shape representations, while the SOS-Net tries to predict more realistic shapes that are consistent with prior medical knowledge.

\begin{figure*}[hbt!]
\centering
    \includegraphics[height=5.5cm]{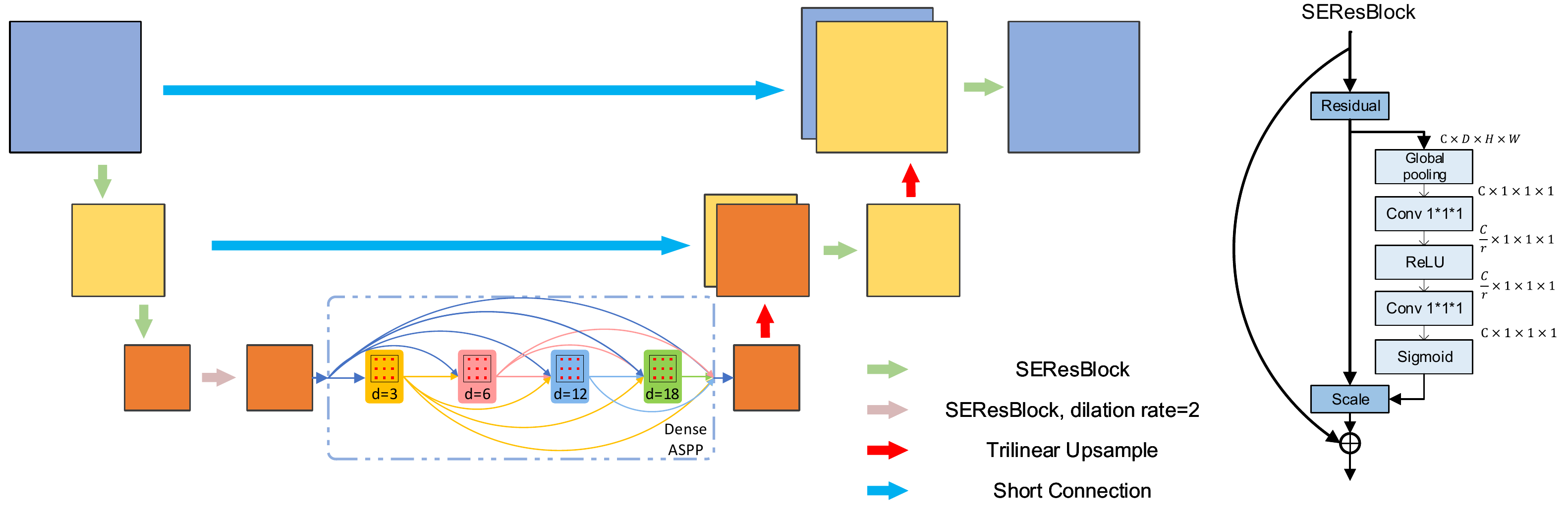}
    \caption{The structure of our S-Net for organ segmentation with multi-scale feature learning. The squares denote feature volumes, and the side length indicates the feature volume size. The \emph{d} in dense ASPP denotes the dilation rate of each convolution kernel.}\label{fig:S-Net}
\end{figure*}

\subsection{Segmentation Network}
\subsubsection {Main Segmentation Network (S-Net)}

U-Net is a commonly used 2D CNN that has made great progress on medical image segmentation compared to conventional methods. Recently, its 3D variants were shown to tackle the segmentation problem of 3D images by better capturing volumetric contextual information. However, vanilla 3D U-Net has poor performance on head-and-neck OAR segmentation. We observe that its inefficacy mainly derives from two aspects. First, U-Net has a symmetric encoder-decoder structure, the encoder embeds multi-scale information into feature maps by four down-sampling operations, while the decoder gradually reconstructs spatial resolution from high-level feature maps by a series of up-sampling or deconvolution operations. However, too much down-sampling leads to the loss of high-resolution information, which would have catastrophic effects on the small organs that only occupy a few voxels. Although high-resolution features are utilized in the shortcut connections between the encoder and decoder, the fusion of low-level and high-level features can only alleviate the problem to some extent. Second, multi-scale features are learned through multiple times of down-sampling, which makes the scale of high-level features fixed and have limited representation capability.

Our proposed S-Net is designed to solve the problem mentioned above. As shown in Fig. \ref{fig:S-Net}, S-Net has a strong backbone, which is a variant of 3D U-Net with residual connections. Squeeze-and-excitation modules \citep{Hu_2018_CVPR} are used for channel-wise attention. The S-Net is built upon the SEResBlock, where the detailed structure is exhibited in Fig. \ref{fig:S-Net}. To reduce information loss and balance between GPU memory usage and segmentation accuracy, the S-Net only performs 2X down-sampling twice. However, such a structure has the disadvantage of limited receptive fields, making it difficult to integrate global image patterns to learn high-level features. Therefore, dilated convolution and densely connected atrous spatial pyramid pooling (DenseASPP) \citep{Yang_2018_CVPR} are also adopted in our S-Net. Densely connected ASPP has the ability to combine arbitrary scales of features via adjusting the dilation rate with better feature reuse. In our model, we use the dilation rates of 3, 6, 12, 18. It should be noted that the dilation is only applied to \emph{x} and \emph{y} axes of the convolution kernels, while the dilation rates along \emph{z} axis are fixed to 1. This is because many small organs only present in several continuous slices. Shortcut connections are also utilized to fuse features of the same scales from the encoder and decoder for learning better features.

\subsubsection {Small-Organ Localization Network (SOL-Net)}

Although the S-Net uses several components to improve the performance, the class imbalance problem between the large and small organs still prevents the network from accurately segmenting small organs. Doctors usually zoom in to finely delineate small organs, which is similar to ROI-pooling operations in two-stage detection or instance segmentation methods.

For OAR localization, the location, orientation and size of OARs are generally consistent among patients, and would not change as general objects in natural images. Therefore, directly regressing the small-organ keypoint locations is more suitable for OAR localization. Inspired by keypoint detection tasks \citep{newell2016stacked}, we propose to design a Small-Organ Localization Network (SOL-Net) to localize the center locations of small organs. As shown in Fig. \ref{fig:overview}, the feature volumes from the last layer of the decoder of our S-Net is used as the input of SOL-Net. The training targets are the small-organ center location heat maps created as 3D Gaussian distributions centered at the small-organ center locations with a standard deviation of 5 voxels, and each small organ has a separate map. The SOL-Net is trained to predict such center-location heat maps with the $L2$ loss. It consists of 1 SEResBlock and a final $1 \times 1 \times 1$ convolution layer with a sigmoid layer to output the small-organ location probability maps. With such location maps, we could further ROI-pool feature volumes from the small organ locations for accurately segmenting them.

\subsubsection{Small-Organ Segmentation Network (SOS-Net)}

Given the center locations of the small organs from the SOL-Net outputs, we further improve the segmentation accuracy by focusing on each small organ's surrounding regions. Specifically, given the location probability map of a small organ, we first identify the voxel with the highest location probability value as the small-organ center location, and ROI-pool a 3D feature volume around it. Considering the localization error, and the sizes of small organs in our collected dataset, the size of ROIs are fixed as $8\times 64\times 64$ ($24\times 64\times 64$mm in physical space) for all small organs. This is different from our original FocusNetv1 \citep{gao2019focusnet}, where the ROI sizes are set as three-time of the OARs' sizes, for implementation consideration, as we add adversarial autoencoder for regularizing small organs' estimated shapes. A Small-Organ Segmentation Network (SOS-Net) is adopted for each small organ's segmentation.

To add more high-resolution information for the segmentation of small organs, multi-scale feature volumes from the last layer of the S-Net decoder as well as the raw input image are ROI-pooled from the small-organ ROI and concatenated together as the input of SOS-Net. Intuitively, the multi-scale feature volumes from S-Net already encode small organs' initial segmentation results and the raw image volumes can help refine the segmentation results. The SOS-Net consists of 2 SEResBlocks and a $1 \times 1 \times 1$ convolution layer with a sigmoid layer to output the binary mask of each small organ. The proposed two-stage localization-refinement strategy for small organ segmentation can simultaneously eliminate information loss from the down-samplings and the class imbalance between large and small organs. To output unified segmentation results of all organs, we overlay the small-organ segmentation results by the SOS-Net into the large-organ segmentation results from the S-Net to obtain the final segmentation map for all organs.

\subsection{Segmentation losses}

In our task, the ratio between the largest and the smallest organ can reach nearly ${500:1}$, which makes the loss dominated by the large number of large-organ voxel samples. In recent literature, focal loss \citep{lin2017focal} and generalized dice loss are two effective loss functions that try to solve the problem of class imbalance. We propose to use a weighted focal loss for multi-class segmentation,
\begin{equation}
\mathcal{L}_{focal}=\sum_{t=0}^{C}-\alpha_t(1-p_t)^\gamma \log(p_t),
\end{equation}
where $C$ is the number of categories, $p_t$ is the probability of class $t$. For training the S-Net, $\alpha_t$ is the weight of each organ, which is inversely proportional to each organ's average size. ${(1-p_t)^\gamma}$ is the modulating factor that weights less on easy samples (voxels) with prediction confidence $p_t$ close to $1$ and weight larger on challenging samples. The focal loss can adaptively down-weight the loss contributed from easy samples, while suppress lightly the contributions of incorrectly classified hard samples. In our experiment, $\gamma$ is empirically set as $2$.

Generalized dice loss is another loss function that directly optimizes towards the evaluation metrics. We adopt the following generalized dice loss,
\begin{equation}
\mathcal{L}_{dice}=\sum_{t=0}^{C}\left(1 - 2\frac{\sum y_tp_t}{\sum y_t + \sum p_t}\right),
\end{equation}
where $y_t$ and $p_t$ denote the ground-truth labels and predicted probabilities of class $t$.
In our experiment, the combination of focal loss and dice loss results in the best segmentation accuracy. The total segmentation loss is therefore defined as
\begin{equation}
\mathcal{L}_{seg}=\mathcal{L}_{focal}+\lambda \mathcal{L}_{dice},
\end{equation}
where $\lambda$ weights the two losses and we empirically set $\lambda=1$ in our experiments.

\subsection{Shape regularization with adversarial autoencoder (AAE)}

\begin{figure}[hbt!]
\centering
    \includegraphics[height=4.3cm]{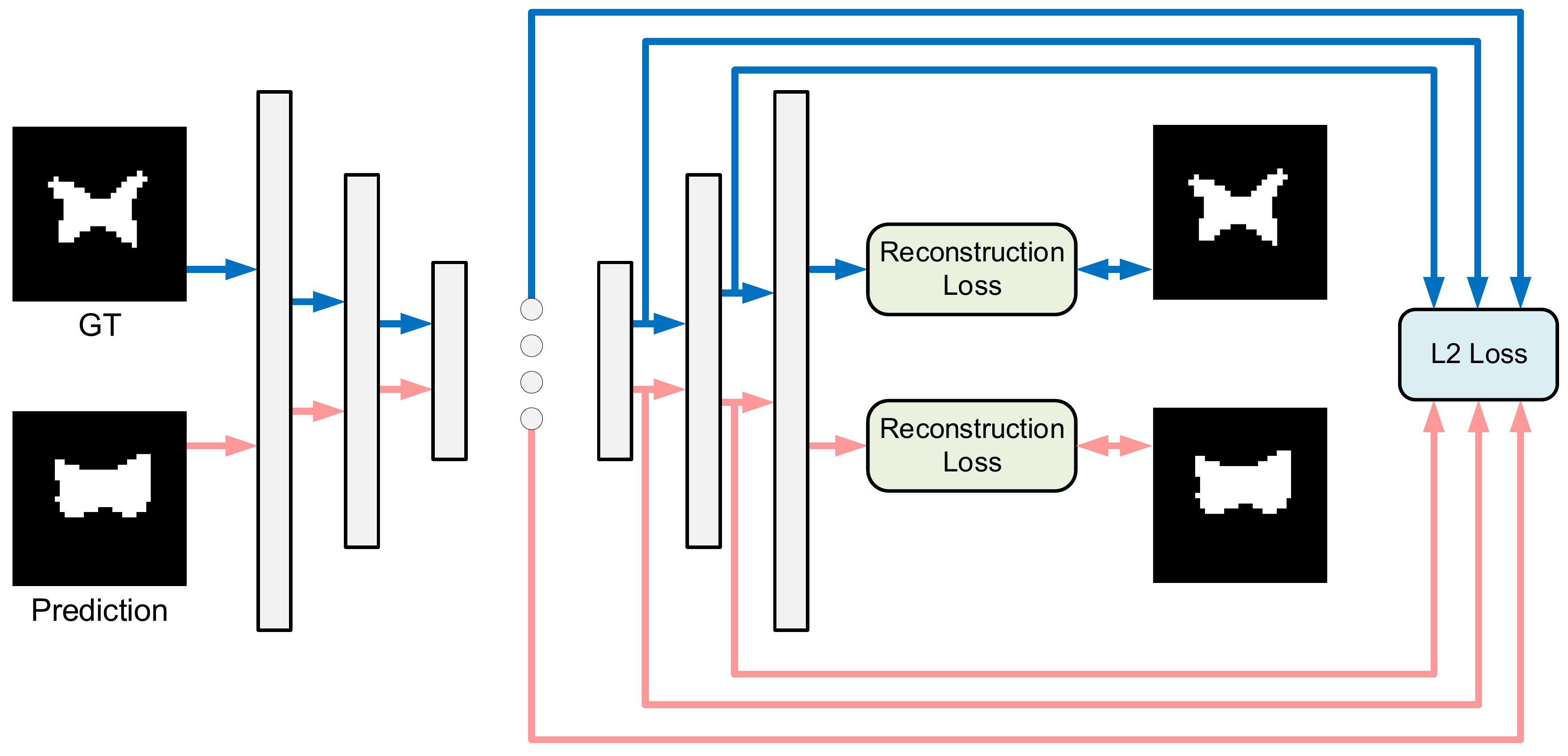} \\
    \caption{The structure of adversarial autoencoder with the shape reconstruction loss and adversarial $L2$ loss.}\label{fig:AAE}
    
\end{figure}

Due to CT's imaging principle, some OARs do not show obvious boundaries in the images, such as the optic chiasm. As OARs are normal organs, they usually have relatively consistent shapes among different patients, incorporating high-order shape constraint to the segmentation network can make the prediction more consistent with prior anatomical knowledge. We propose a novel adversarial autoencoder (AAE) to introduce shape regularization into the training of our segmentation framework. To the best of our knowledge, this is the first segmentation method that leverages both the autoencoder and adversarial learning for constraining segmentation masks.

A good design of shape regularization term should have the following two characteristics. First, it needs to be capable of representing shapes in a differentiable way, so that the segmentation network can be trained with back-propagation for shape regularization; Second, it should be able to distinguish subtle differences between shapes, such that as the shape predicted by the segmentation network gets closer to the ground-truth shapes, it still gives a correct penalty.

The shapes represented by label maps are highly structured and high dimensional, making it extremely challenging to measure the similarity between two shapes in such high dimensional space. The high-dimensional shape usually lies in a lower-dimensional shape manifold \citep{wang2014generalized, oktay2017anatomically}, where each shape would be mapped to a lower-dimensional point (vector) in the subspace.  If a shape manifold is successfully discovered, starting from a point on the manifold (corresponding to one specific shape), we can traverse along with different directions on the manifold. The corresponding shape would change smoothly and continuously at the semantic level. Therefore, we measure the similarity between two shapes in a low-dimensional shape manifold. We use a shape autoencoder, which is a neural network trained to reconstruct the input organ shape as much as possible (see Fig. \ref{fig:AAE}). The bottleneck structure enables the autoencoder to transform the input shape into a latent code that captures its salient features while discarding irrelevant features. Therefore, if the autoencoder can well reconstruct the input shape, the latent space is a good approximation of low-dimensional shape manifold \citep{lei2020geometric, zhu2016generative}. Moreover, the autoencoder is differentiable and can regularize estimated organ shapes by minimizing the distance between the predicted organ shape and ground-truth shape in the latent space.

For the second characteristic, the accurate measurement of the similarity between the predicted organ shape and the ground-truth organ shape is crucial. Therefore, we introduce an adversarial training scheme for training the adversarial autoencoder. The autoencoder is trained with both the predicted shape from the small organ segmentation branch and the corresponding ground-truth shapes. It has two loss terms, where the first one is to reconstruct the input shapes for learning shape representations by minimizing the conventional reconstruction loss,
\begin{equation}
    \mathcal{L}_{rec}=|D(y)-y|^2_2 + |D(G(x))-G(x)|^2_2
\end{equation}
where $x$ is the input image, $y$ is its corresponding ground-truth label, $G$ is the segmentation network, $G(x)$ is the predicted binary organ mask from SOS-Net given the input image, $D(y)$ and $D(G(x))$ are the reconstruction results of AAE, given the ground truth label $y$ and predicted organ mask $G(x)$.



The other adversarial loss term tries to distinguish the latent codes of predicted shapes and ground-truth shapes by maximizing their distance in the low-dimensional manifold. In this way, we enforce the autoencoder to better encode the two types of shapes and capture their subtle differences, while the segmentation network is encouraged to fool the autoencoder to being unable to capture the subtle differences.
Therefore, the proposed adversarial shape loss is formulated as
\begin{equation}
   \mathop{min}\limits_{G}\mathop{max}\limits_{D}\ \mathcal{L}_{shape}= \mathbb{E}_{x\sim p_{data},\ y\sim p_{gt}}\|D_{latent}(y)- D_{latent}(G(x))\|^2_2,
\end{equation}
where $D_{latent}(y)$ or $D_{latent}(G(x))$ is the latent code and multi-scale decoder features of the ground-truth organ shape $y$ and the predicted organ shape $G(x)$, as illustrated in Fig. \ref{fig:AAE}. Intuitively, the segmentation network $G$ and the AAE $D$ play a min-max game. The segmentation network $G$ tries to predict masks that are consistent with shape prior to minimize the distance in low-dimensional shape manifold, while the AAE $D$ tries to learn better encoding to maximize the distance.

The overall objective function for segmentation network and with the shape regularization term from the adversarial autoencoder is therefore defined as
\begin{equation}
    \mathop{min}\limits_{G}\mathop{max}\limits_{D}\ \mathcal{L}_{seg}+\lambda\mathcal{L}_{shape}.
\end{equation}
In practice, the segmentation network $G$ and the adversarial autoencoder $D$ are trained in an alternative optimization scheme. The $G$ is first optimized by fixing $D$ and minimizing the following loss
\begin{align}
	\mathop{min}\limits_{G}\ L_G = \mathcal{L}_{seg}+\lambda_1\mathcal{L}_{shape},
	\label{l_g}
\end{align}
where $\lambda_1$ is empirically set as 5 to balance the two terms. Optimizing $L_G$ would encourage the segmentation network $G$ to output organ shapes that are consistent with the ground-truth shapes.
The $D$ is then optimized when $G$ is fixed by minimizing the following loss,
\begin{align}
	\mathop{max}\limits_{D}\ L_D = -\mathcal{L}_{rec}+\lambda_2\mathcal{L}_{shape},
	\label{l_d}
\end{align}
where $\lambda_2$ is set as 0.001 in our experiment as larger $\lambda_2$ may cause unstable training of the proposed AAE. During the training of segmentation network $G$, the estimated organ shapes by $G$ will gradually become closer to the ground-truth organ shapes, therefore, if the encoded latent codes are not distinguishable between the estimated organ shapes and the ground-truth organ shapes, the autoencoder may hardly provide effective supervision for the segmentation network $G$. Therefore, maximizing the distance between $D_{latent}(G(x))$ and $D_{latent}(y)$ encourages the autoencoder to encode subtle difference between them. 
Eqs. \eqref{l_g} and \eqref{l_d} are optimized alternatively to gradually improve both the segmentation network $G$ and the adversarial autoencoder $D$.

{\bf Discussions on segmentation with autoencoder or adversarial learning.} Previous segmentation methods have also explored using autoencoder in segmentation. \citet{oktay2017anatomically} and \citet{tong2018fully} only trained the autoencoder with ground-truth organ shapes. The segmentation network is then trained to minimize the distance between the latent features from autoencoder of predicted shapes and ground truth shapes, where the parameters of autoencoder are fixed. However, since the autoencoder has not been trained with the predicted organ shapes by the segmentation network ever, when the estimated organ shapes are close to the ground-truth shapes, the latent codes of autoencoder alone cannot distinguish these two, thus the autoencoder cannot provide effective guidance to further regularize the segmentation network. Moreover, in \citet{tong2018fully}, the autoencoder is trained with whole images. The autoencoder training also faces the extremely unbalanced data, thus making the training of small organs' shape representations ineffective. In our approach, we adopt the autoencoder in the small-organ branch, thus avoid the imbalanced class problem.

Several previous works \citep{yang2017automatic, han2018spine} adopted adversarial training in segmentation networks using the objective function in the following way,
\begin{equation}
    \mathop{min}\limits_{G}\mathop{max}\limits_{D}\
        \mathbb{E}_{y\sim P_{gt}}[\log(D(y)]+\mathbb{E}_{x\sim P_{data}}[\log(1-D(G(x))],
\end{equation}
where $x$ is the input image, $y$ is its corresponding ground-truth shape, $G$ is the segmentation network, $G(x)$ is the predicted organ shape, $D$ is the discriminator network that tries to distinguish its input to be real or fake (i.e. ground truth label or network predicted mask). However, $D$ was previously designed as a classifier while we adopted the shape autoencoder for the min-max game.

\begin{figure*}[hbt!]
\centering
    \includegraphics[height=20cm]{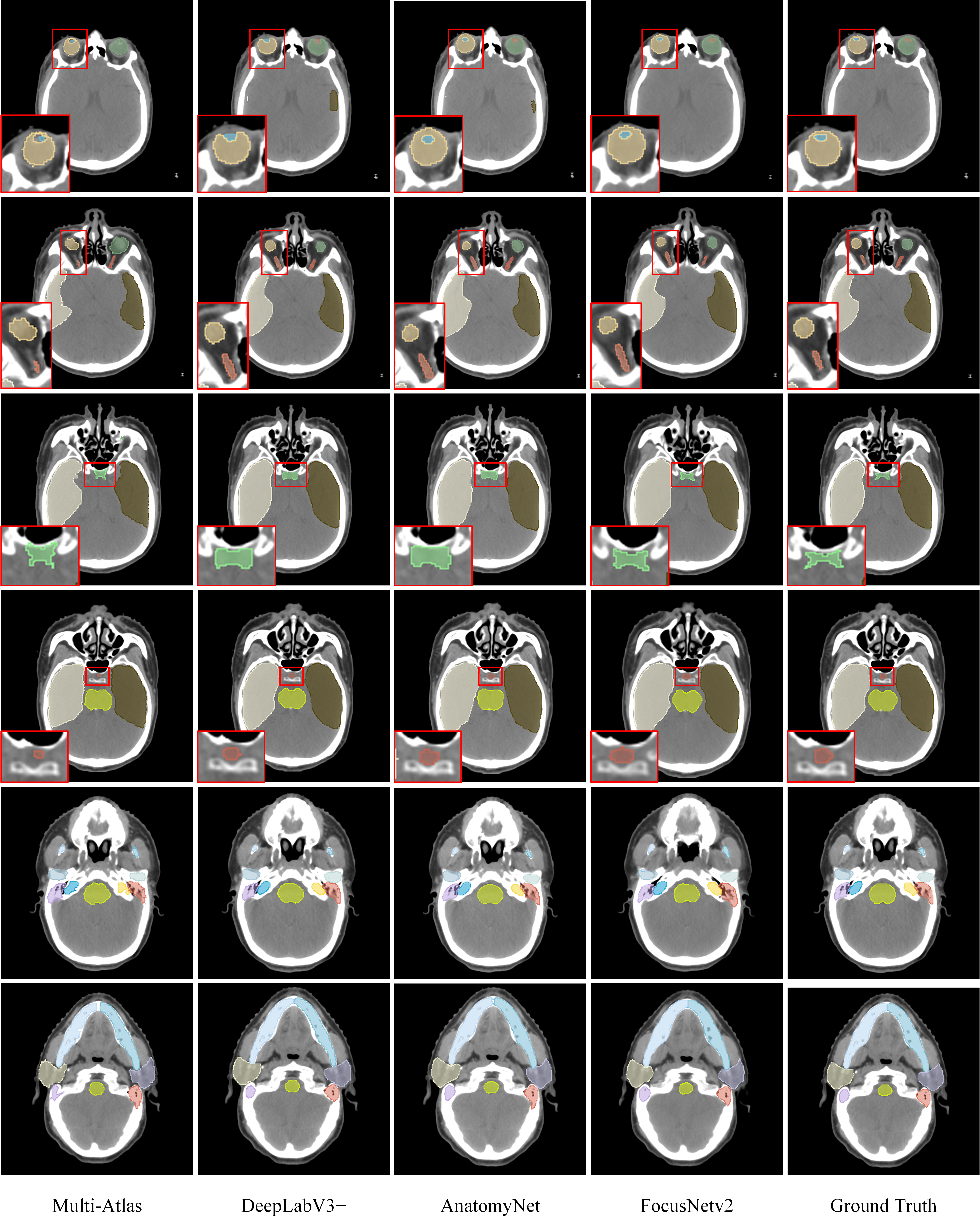}
    \caption{Visualization of results by compared methods on our dataset.} \label{fig:visual}
\end{figure*}

\begin{table*}[htb!]

\begin{center}
\setlength{\tabcolsep}{5mm}
\begin{tabular}{lccccc}
\toprule
OAR                 &Multi-Atlas                &DeepLabv3+             &AnatomyNet                  &FocusNetv1          &FocusNetv2 \\
\midrule
Eye L               &78.17$\pm$11.28    &88.78$\pm$2.34         &88.27$\pm$2.59             &89.28$\pm$1.95            &\textbf{89.68$\pm$2.22} \\      
Eye R               &78.62$\pm$10.15    &87.97$\pm$2.79         &87.56$\pm$2.40             &88.95$\pm$2.34          &\textbf{89.47$\pm$2.12}      \\
\rowcolor{mygray}
Lens L              &30.45$\pm$18.77    &72.86$\pm$7.21         &72.72$\pm$9.86              &78.06$\pm$8.37          &\textbf{81.91$\pm$8.36} \\
\rowcolor{mygray}
Lens R              &28.44$\pm$16.32    &74.82$\pm$68.21        &74.92$\pm$8.68             &78.73$\pm$7.56          &\textbf{82.47$\pm$6.46} \\
\rowcolor{mygray}
Opt Nerve L         &41.23$\pm$16.32    &68.21$\pm$9.88          &63.91$\pm$11.27           &68.76$\pm$9.98          &\textbf{71.33$\pm$9.23} \\
\rowcolor{mygray}
Opt Nerve R         &44.15$\pm$17.45    &69.90$\pm$9.31          &65.41$\pm$10.59            &73.32$\pm$8.84          &\textbf{75.25$\pm$7.59} \\
\rowcolor{mygray}
Opt Chiasm          &40.07$\pm$15.66    &57.70$\pm$14.72          &54.89$\pm$14.72          &61.15$\pm$12.44          &\textbf{61.22$\pm$13.34} \\
\rowcolor{mygray}
Pituitary           &41.13$\pm$18.56    &67.21$\pm$12.05          &66.13$\pm$10.88          &68.78$\pm$12.55          &\textbf{72.19$\pm$11.88} \\
Brain Stem          &86.14$\pm$4.61     &88.91$\pm$3.06          &88.51$\pm$3.26            &\textbf{89.26$\pm$3.17} &89.09$\pm$3.47          \\
Temporal Lobes L    &86.69$\pm$5.51     &87.52$\pm$4.00          &87.31$\pm$4.89            &\textbf{87.89$\pm$3.94} &87.73$\pm$3.98          \\
Temporal Lobes R    &87.39$\pm$5.16     &87.99$\pm$4.19          &87.53$\pm$3.91            &88.22$\pm$3.85          &\textbf{88.30$\pm$4.30}  \\
Spinal Cord         &78.22$\pm$5.7      &81.99$\pm$5.59          &82.79$\pm$5.26             &82.60$\pm$5.32          &\textbf{83.09$\pm$5.04} \\
Parotid Gland L     &80.55$\pm$7.91     &82.99$\pm$5.56          &83.89$\pm$5.76            &\textbf{85.05$\pm$5.47} &84.58$\pm$6.06          \\
Parotid Gland R     &83.45$\pm$5.07     &85.91$\pm$2.96          &85.88$\pm$3.37           &86.87$\pm$3.82          &\textbf{87.01$\pm$3.80}          \\
Inner Ear L         &85.04$\pm$6.75     &84.88$\pm$5.08          &\textbf{86.79$\pm$5.43}   &84.87$\pm$5.48          &86.42$\pm$4.77          \\
Inner Ear R         &81.91$\pm$6.57     &84.48$\pm$5.44          &86.66$\pm$6.34            &85.04$\pm$5.96          &\textbf{85.51$\pm$5.90} \\
Mid Ear L           &82.44$\pm$7.74     &84.27$\pm$5.97          &85.71$\pm$6.21            &\textbf{86.10$\pm$5.63} &85.73$\pm$5.90          \\
Mid Ear R           &79.23$\pm$11.4     &82.39$\pm$7.92          &84.02$\pm$8.18            &84.12$\pm$8.14          &\textbf{84.32$\pm$8.23} \\
TMJ L               &74.98$\pm$10.86    &76.22$\pm$10.13         &\textbf{77.76$\pm$10.37}  &76.43$\pm$10.92       &76.36$\pm$9.98          \\
TMJ R               &78.15$\pm$9.69     &\textbf{79.12$\pm$9.58} &79.11$\pm$10.69           &78.65$\pm$12.31          &78.91$\pm$10.07          \\
Mandible L          &89.28$\pm$3.35     &90.70$\pm$1.56          &92.00$\pm$1.61           &92.01$\pm$1.57          &\textbf{92.44$\pm$1.63}          \\
Mandible R          &89.61$\pm$3.44     &90.08$\pm$2.18          &91.84$\pm$1.46    &92.21$\pm$1.53                 &\textbf{92.54$\pm$1.63}          \\
\midrule
Average             &70.24              &80.70                  &80.62                      &82.10          &\textbf{82.98} \\
Small Organ Average &37.57              &68.45                  &66.33                      &71.47          &\textbf{74.04} \\

\bottomrule
\end{tabular}
\end{center}
\caption{Dice score coefficient (\%) of results by different compared methods on our collected dataset. Shaded rows indicate that those OARs are treated as small organs.}\label{tab:result_dsc}
\end{table*}

\section{Experiments}

\subsection{Datasets}

The proposed FocusNetv2 was evaluated on two datasets of HaN CT images. The first dataset is a self-collected dataset, denoted as our dataset. Our dataset consists of 1,164 collected CT scans of patients with nasopharyngeal carcinoma. 22 OARs to be considered in HaN radiotherapy treatment planning are delineated in each scan, including (left and right) eyes, (left and right) lens, (left and right) optic nerves, optic chiasm, pituitary, brain stem, (left and right) temporal lobes, spinal cord, (left and right) parotid glands, (left and right) inner ears, (left and right) middle ears, (left and right) temporomandibular joints, and (left and right) mandible. Ground truth annotations of each case were provided by senior doctors with hundreds of cases of annotation experience, with each structure being segmented by the same annotator and reviewed by the other one. The left and right lens, left and right optic nerves, optic chiasm, pituitary are defined as small organs due to their small volume and complex anatomical structures. The CT scans have anisotropic voxel spacing ranging from 0.78mm to 1.25mm and inter-slice thickness ranging from 2.7mm to 3.5mm. All scans are resampled to $1\times 1\times 3$ mm for further processing. We randomly shuffle our dataset and select 1,044 samples for training and 120 samples for testing.

For comparison with state-of-the-art methods on HaN OAR segmentation, we evaluate the proposed FocusNetv2 on a public dataset, the \emph{MICCAI Head and Neck Auto Segmentation Challenge 2015} dataset (denoted as MICCAI'15 dataset). This dataset is also known as a Public Domain Database for Computational Anatomy (PDDCA) provided and maintained by Harvard Medical School. The dataset includes multiple image studies from patients with stage III or IV squamous cell carcinoma of the oropharynx, hypopharynx or larynx. It consists of 38 CT scans for training and 10 scans for testing, and has 9 organ annotations: brain stem, mandible, optic chiasm, (left and right) optic nerves, (left and right) parotid glands and (left and right) submandibular glands, where optic chiasm, left and right optic nerves are defined as small organs. The delineation of structures is based on the protocol described by Radiation Therapy Oncology Group (RTOG). We resample all scans to have a voxel size of $1 \times 1 \times 2.5$ mm to train our FocusNetv2, while for a fair comparison with other methods, we resampled the predicted segmentation labels by our proposed method back into the original spacing and then calculate the evaluation metrics.

\subsection{Implementation details}

Our method is implemented with PyTorch and trained on NVIDIA TITAN Xp GPUs. The segmentation networks are trained from scratch with initial weights sampled from a standard Gaussian distribution. We first train the S-Net, and then train the SOL-Net while fixing the trained parameters of S-Net. The SOS-Net is trained afterward, and is updated with the adversarial autoencoder in an alternative way. At last, we finetune the whole network for joint optimization. We use the ADAM optimizer to train the network with a learning rate 0.0005. The batch size is set as 1. For the adversarial autoencoder, it is pretrained with ground truth labels in the dataset to stabilize the adversarial training process. 

 The original CT image size was around $n\times 512\times 512$ with $n$ being the number of slices. Since the majority of each CT image is the background, they are centrally cropped to $n\times 240\times 240$. 
 It would be better to use whole image volumes to train the network. However, due to GPU memory limitation, we randomly crop 40-slice chunks along the $z$-axis from CT images for each iteration when training the S-Net and the SOL-Net. One problem raised by the sliding slices strategy is that the cropping process might destroy organs' shape for training AAE and SOS-Net. As the small organs in both datasets are mainly lens, optic nerves, optic chiasm, and pituitary. They lie in only a few adjacent slices along the $z$-axis, which can be fully included in a 40-slice cube with a large margin. Therefore, we adopt a sampling strategy when training AAE and SOS-Net. Although we sample cubes along $z$-axis with random translation, we always ensure that the cube contains all small organs with some margin. Therefore, the shapes of small organs are complete. The CT scans are cropped every 40 slices with a stride of 40 along the $z$-axis, i.e., no overlap between the crops. We then stack the segmentation result of each 40-slice crop together to obtain the final prediction. Random affine transformations (translation within 40 pixels in x and y axes, rotation within 10 degrees, and scale from 0.7 to 1.3 times) are used for data augmentation during training.

\subsection{Evaluation metrics}

We use two evaluation metrics in this study. Dice score coefficient (DSC) measures the degree of overlap between the predicted segmentation and the ground truth segmentation with the formula, $DSC(X,Y)=\frac{2\mid X\cap Y\mid}{\mid X\mid + \mid Y\mid}$, where $X$ and $Y$ represents the voxel sets of prediction and ground-truth respectively.
95\% Hausdorff Distance (95HD) is a variant of Hausdorff distance, which measures the largest distance from points in $X$ to its nearest neighbors in $Y$. The HD is calculated as the average of two directions, $HD = (d_H(X, Y) + d_H(Y, X))/2$. The 95\% Hausdorff Distance can mitigate the sensitivity of outliers by calculating the 95\% largest distances.

\begin{table*}[t!]

\begin{center}
\setlength{\tabcolsep}{5mm}
\begin{tabular}{lccccc}
\toprule
OAR                 &Multi-Atlas             &DeepLabv3+          &AnatomyNet                 &FocusNetv1       &FocusNetv2 \\
\midrule
Eye L               &5.18$\pm$6.45      &1.67$\pm$0.71      &1.74$\pm$0.62           &1.64$\pm$0.80          &\textbf{1.51$\pm$0.63} \\      
Eye R               &5.22$\pm$6.00      &1.84$\pm$0.70      &2.03$\pm$0.71           &1.81$\pm$0.73          &\textbf{1.56$\pm$0.60}    \\
\rowcolor{mygray}
Lens L              &5.68$\pm$3.61      &1.99$\pm$0.88      &2.04$\pm$0.91           &1.57$\pm$0.77          &\textbf{1.41$\pm$0.68} \\
\rowcolor{mygray}
Lens R              &5.34$\pm$2.52      &1.94$\pm$0.84      &1.76$\pm$0.83           &1.73$\pm$1.05          &\textbf{1.59$\pm$1.07} \\
\rowcolor{mygray}
Opt Nerve L         &4.28$\pm$1.58      &2.77$\pm$0.84      &3.21$\pm$0.96           &2.64$\pm$0.81          &\textbf{2.52$\pm$0.89} \\
\rowcolor{mygray}
Opt Nerve R         &4.12$\pm$1.44      &2.89$\pm$0.65      &3.14$\pm$0.87           &2.61$\pm$0.89          &\textbf{2.55$\pm$0.89} \\
\rowcolor{mygray}
Opt Chiasm          &5.16$\pm$1.58      &3.77$\pm$0.98      &3.83$\pm$0.95           &\textbf{3.47$\pm$0.80} &3.58$\pm$0.93 \\
\rowcolor{mygray}
Pituitary           &4.03$\pm$1.40      &2.49$\pm$0.84      &2.62$\pm$0.86           &2.36$\pm$0.87          &\textbf{2.16$\pm$0.94} \\
Brain Stem          &2.61$\pm$1.37      &2.09$\pm$1.17      &2.15$\pm$1.16           &2.05$\pm$1.18          &\textbf{1.93$\pm$0.96}          \\
Temporal Lobes L    &4.44$\pm$3.43      &4.01$\pm$3.21      &4.14$\pm$3.16               &\textbf{3.95$\pm$3.37} &3.99$\pm$3.19          \\
Temporal Lobes R    &4.14$\pm$3.22      &\textbf{3.78$\pm$3.00} &4.05$\pm$2.71           &3.87$\pm$2.96          &3.85$\pm$2.95  \\
Spinal Cord         &6.26$\pm$7.72      &1.86$\pm$0.89      &1.75$\pm$0.92              &1.75$\pm$0.89          &\textbf{1.70$\pm$0.68} \\
Parotid Gland L     &4.09$\pm$3.09      &3.68$\pm$2.84      &3.43$\pm$2.35           &\textbf{3.26$\pm$2.54} &3.30$\pm$2.47          \\
Parotid Gland R     &3.44$\pm$1.63      &2.93$\pm$1.40      &2.76$\pm$1.05           &\textbf{2.58$\pm$1.33} &2.63$\pm$1.28          \\
Inner Ear L         &2.23$\pm$0.94      &2.03$\pm$0.88      &\textbf{1.81$\pm$0.93}  &2.09$\pm$0.86          &1.99$\pm$0.85          \\
Inner Ear R         &2.65$\pm$0.86      &2.01$\pm$0.81      &\textbf{1.82$\pm$0.95}  &2.11$\pm$0.87          &2.09$\pm$0.92 \\
Mid Ear L           &3.09$\pm$1.71      &3.09$\pm$1.19      &2.69$\pm$1.35           &\textbf{2.53$\pm$1.2}  &2.59$\pm$1.31         \\
Mid Ear R           &4.44$\pm$3.71      &3.95$\pm$2.49      &3.31$\pm$2.12           &\textbf{3.12$\pm$1.81} &3.37$\pm$2.74 \\
TMJ L               &3.29$\pm$1.55      &3.10$\pm$1.25      &3.02$\pm$1.53           &\textbf{3.07$\pm$1.22} &3.15$\pm$1.29          \\
TMJ R               &2.74$\pm$0.89      &2.74$\pm$1.00      &2.67$\pm$1.18           &\textbf{2.56$\pm$1.15} &2.66$\pm$1.07          \\
Mandible L          &2.56$\pm$2.25      &1.59$\pm$0.67      &1.38$\pm$0.70           &1.42$\pm$0.67          &\textbf{1.31$\pm$0.66}  \\
Mandible R          &2.47$\pm$2.52      &1.54$\pm$0.64      &1.28$\pm$0.64           &\textbf{1.25$\pm$0.55}  &1.26$\pm$1.26  \\
\midrule
Average             &3.98           &2.63                   &2.58                    &2.43                  &\textbf{2.40} \\
Small Organ Average &4.76           &2.64                   &2.76                    &2.39                  &\textbf{2.30} \\

\bottomrule
\end{tabular}
\end{center}
\caption{95th percentile HD (mm) of results by different compared methods on our collected dataset. Shaded rows indicate that those OARs are treated as small organs.}\label{tab:result_95hd}
\end{table*}

\begin{table*}[tb!]
\footnotesize
\begin{center}
\setlength{\tabcolsep}{2mm}
\begin{tabular}{lcc>{\columncolor{mygray}}c>{\columncolor{mygray}}c>{\columncolor{mygray}}ccccccc}

\toprule
Study                       &Brain              &Mandible           &Optic           &Optic         &Optic      &Parotid        &parotid        &SMG         &SMG               &AVG   &Small\\
                            &Stem               &                   &Chiasm          &Nerve L       &Nerve R    &L              &R              &L           & R                &       &AVG\\
\midrule
\citet{raudaschl2017evaluation}    &88.0               &93.0               &55.0           &62.0           &62.0       &84.0           &84.0           &78.0           &78.0    &76.0   &59.7\\
\citet{ren2018interleaved}  &-                  &-                  &58.0$\pm$17.0  &72.0$\pm$8.0   &70.0$\pm$9.0 &-            &-              &-              &-              &-      &-\\
\citet{wang2017hierarchical}&\textbf{90.0$\pm$4.0}  &94.0$\pm$1.0   &-              &-            &-          &83.0$\pm$6.0  &83.0$\pm$6.0    &-              &-              &-        &-\\
\citet{zhu2019anatomynet}   &86.7$\pm$2.0       &92.5$\pm$2.0       &53.2$\pm$15.0  &72.1$\pm$6.0   &70.6$\pm$10.0 &88.1$\pm$2.0 &87.3$\pm$4.0  &81.4$\pm$4.0   &81.3$\pm$4.0   &79.3   &65.3\\
\citet{tong2018fully}       &87.0$\pm$3.0       &93.7$\pm$1.2       &58.4$\pm$10.3  &65.3$\pm$5.8   &68.9$\pm$4.7  &83.5$\pm$2.3  &83.2$\pm$1.4 &75.5$\pm$6.5   &81.3$\pm$6.5   &77.4   &64.2\\
\citet{tang2019clinically}  &87.5$\pm$2.5       &\textbf{95.0$\pm$0.8} &61.5$\pm$10.2 &74.8$\pm$7.1 &72.3$\pm$5.9  &88.7$\pm$1.9  &87.5$\pm$5.0 &82.3$\pm$5.2   &81.5$\pm$4.5   &81.2   &69.5\\
FocusNetv2           &88.2$\pm$2.5       &94.7$\pm$1.1       &\textbf{71.3$\pm$17.0} &\textbf{79.0$\pm$7.5} &\textbf{81.7$\pm$7.3} &\textbf{89.8$\pm$1.6} &\textbf{88.1$\pm$4.2} &\textbf{84.0$\pm$4.6}   &\textbf{83.8$\pm$4.1}    &\textbf{84.5} &\textbf{77.3}\\

\bottomrule
\end{tabular}
\end{center}
\caption{Dice score coefficient (\%) comparison with previous published result on MICCAI2015 dataset, the columns in gray are small organs.}\label{tab:MICCAI_dsc}
\end{table*}

\subsection {Experiments on our collected dataset}

We compare our proposed method with a multi-atlas based method, where Symmetric Normalization (SyN) \citep{avants2008symmetric} is used as the registration method, a 3D variant of DeepLabv3+ \citep{chen2018encoder} and a state-of-the-art deep learning method in HaN OAR segmentation named AnatomyNet \citep{zhu2019anatomynet}.

For the multi-atlas-based method, we randomly select 9 CT scans from the training set as the atlas due to the tie and computational resouce limitation. Symmetric Normalization (SyN) \citep{avants2008symmetric} with its implementation in ANTs software package is used to recover the optimal affine matrix and deformable transformation field between the CT to be segmented and each atlas. 9 label maps are obtained by applying the transformation fields to atlas labels, then the final prediction is obtained by voting.
DeepLabv3+ \citep{chen2018encoder} is a well-known segmentation framework originally designed for 2D semantic segmentation. It uses the spatial pyramid pooling and dilated convolution and achieves state-of-the-art performance on natural image segmentation. We extend their network structure to 3D for volumetric segmentation. It was randomly initialized and trained using the same loss function as our proposed FocusNetv2. AnatomyNet \citep{zhu2019anatomynet} is designed for fast segmentation on whole CT images and has good performance compared with traditional atlas-based methods.

\subsubsection{Quantitative comparison}
Comparative results are shown in Tables \ref{tab:result_dsc} and \ref{tab:result_95hd}. The conventional multi-atlas based method SyN has decent performance on large organs, especially for those with high contrast with surrounding regions, such as mandible. Nevertheless, it results in undesirable segmentation results for organs of small sizes. Deep learning based methods have overwhelming advantages in these circumstances, because small organs have more complex anatomical structures, the multi-atlas based method SyN has limited capability of dealing with complicated and diverse anatomy variations. Among deep learning based methods, even without adversarial autoencoder, our FocusNetv1 performs better in most organs. This is because that our specially designed two-stage framework significantly reduces the extremely unbalanced ratio between background, large organs and small organs. Each small-organ branch can focus on the segmentation of specified organs, where high-resolution detailed information is incorporated for detailed refinement. After incorporating the adversarial shape loss from the proposed adversarial autoencoder, the segmentation accuracy of small organs could be further improved with large margins. Compared with other approaches, our FocusNet has the best Dice scores in 19 out of 22 organs, and has the best 95HD scores in 19 out of 22 organs. In terms of accuracies of small organs, our FocusNetv2 has an improvement of 5.59\% in Dice score compared with other deep learning based methods.

\subsubsection{Qualitative comparison}
As illustrated in Fig. \ref{fig:visual}, in the first two rows, the multi-atlas based method SyN misses the left lens, and does not perform well on segmenting optic nerves, which is consistent with quantitative results. There are errors in the region where eyeballs are attached to the lens by DeepLabv3+. It might be because DeepLabv3+ processes the images in a low resolution and then up-sample to the origin resolution for final prediction, which results in the loss of detailed information. In the third row, for the optic chiasm, although the result of multi-atlas based method is not accurate enough, its shape conforms to prior medical knowledge. DeepLabv3+ and AnatomyNet are more inclined to predict a smooth shape in its results, which is a common problem of state-of-the-art segmentation deep learning networks. This is because the combination of the low contrast of anatomy contour and the extremely small volume of organs makes the training of small organs ineffective. Moreover, per-pixel losses cannot penalize high-level shape errors.  By incorporating the proposed small-organ segmentation branch and adversarial shape autoencoder, our FocusNetv2 generates satisfactory X-shape segmentation masks. This proves that our FocusNetv2 successfully introduces shape constraints into the deep learning framework. It can achieve satisfactory segmentation accuracy and make the segmentation results consistent with prior medical knowledge at the same time.

\subsubsection{Processing time}

\begin{table}[htb!]

\begin{center}
\setlength{\tabcolsep}{1.7mm}
\begin{tabular}{lcccc}
\toprule
Method      &DeepLabv3+	    &AnatomyNet     &S-Net	&FocusNetv2      \\
\midrule
Time (s)    &4.52           &2.28		    &3.33	&4.36\\
\bottomrule
\end{tabular}
\end{center}
\caption{Average inference time of different deep learning models on one CT scan on a NVIDIA TITAN Xp GPU.}\label{tab:process_time}
\end{table}

The processing time of deep learning methods are presented in Table \ref{tab:process_time}, all methods are measured using the same computing platform and an NVIDIA TITAN Xp GPU. Our backbone network S-Net takes 3.33s on average to process one CT scan. After adding SOL-Net and SOS-Net, our FocusNetv2 takes 4.36s, which is still faster than DeepLabv3+, but with much higher segmentation accuracy. Our method consumes more computing resources than AnatomyNet, but less than DeepLabv3+. Considering that radiotherapy treatment planning generally takes several hours and is not a time-sensitive task, our method can achieve optimal performance in a reasonable time.

We further test the doctor's average delineation time with and without using our algorithm's results. It generally takes about one hour for human doctors to delineate 22 OARs in a head and neck CT scan.  If our algorithm's results are used for assistance, the doctors only need to make small modifications on the automatic delineation results for most patients. The entire delineation time only takes 20-30 minutes, which is 1/3 to 1/2 shorter than before. It could dramatically improve the efficiency of radiologists.

\subsection{Experiments on MICCAI'15 dataset}

\begin{table}[htb!]
\footnotesize
\begin{center}
\setlength{\tabcolsep}{0.5mm}
\begin{tabular}{lcccc}
\toprule
Organ                  &\citet{ren2018interleaved}     &\citet{zhu2019anatomynet} &\citet{tong2018fully}  &FocusNetv2    \\
                       
\midrule
Brain Stem              &-                       &6.42$\pm$2.38    &4.01$\pm$0.93                &\textbf{2.32$\pm$0.70}\\
Optic Chiasm            &2.81$\pm$1.56           &5.76$\pm$2.49    &\textbf{2.17$\pm$1.04}     &2.25$\pm$0.85      \\
Mandible                &-                       &6.28$\pm$2.21    &1.50$\pm$0.32    &\textbf{1.08$\pm$0.45}     \\
Optic Ner. L            &2.33$\pm$0.84           &4.85$\pm$2.32    &2.52$\pm$1.04    &\textbf{1.92$\pm$0.80}\\
Optic Ner. R            &\textbf{2.13$\pm$0.96}  &4.77$\pm$4.27    &2.90$\pm$1.88    &2.17$\pm$0.74\\
Parotid L               &-                       &9.31$\pm$3.32    &3.97$\pm$2.15    &\textbf{1.81$\pm$0.43}\\
Parotid R               &-                       &10.0$8\pm$5.09   &4.20$\pm$1.27    &\textbf{2.43$\pm$2.00}\\
Submand. L              &-                       &7.01$\pm$4.44    &5.59$\pm$5.59    &\textbf{2.84$\pm$1.20}\\
Submand. R              &-                       &6.02$\pm$1.78    &4.84$\pm$4.84    &\textbf{2.74$\pm$1.25}\\
Average                 &-                       &6.72             &3.52             &\textbf{2.17}\\

\bottomrule
\end{tabular}
\end{center}
\caption{95HD score (mm) of results by different compared methods on MICCAI'15 dataset.}\label{tab:MICCAI_95HD}
\end{table}

\begin{figure}[hbt!]
\centering
    \includegraphics[height=13cm]{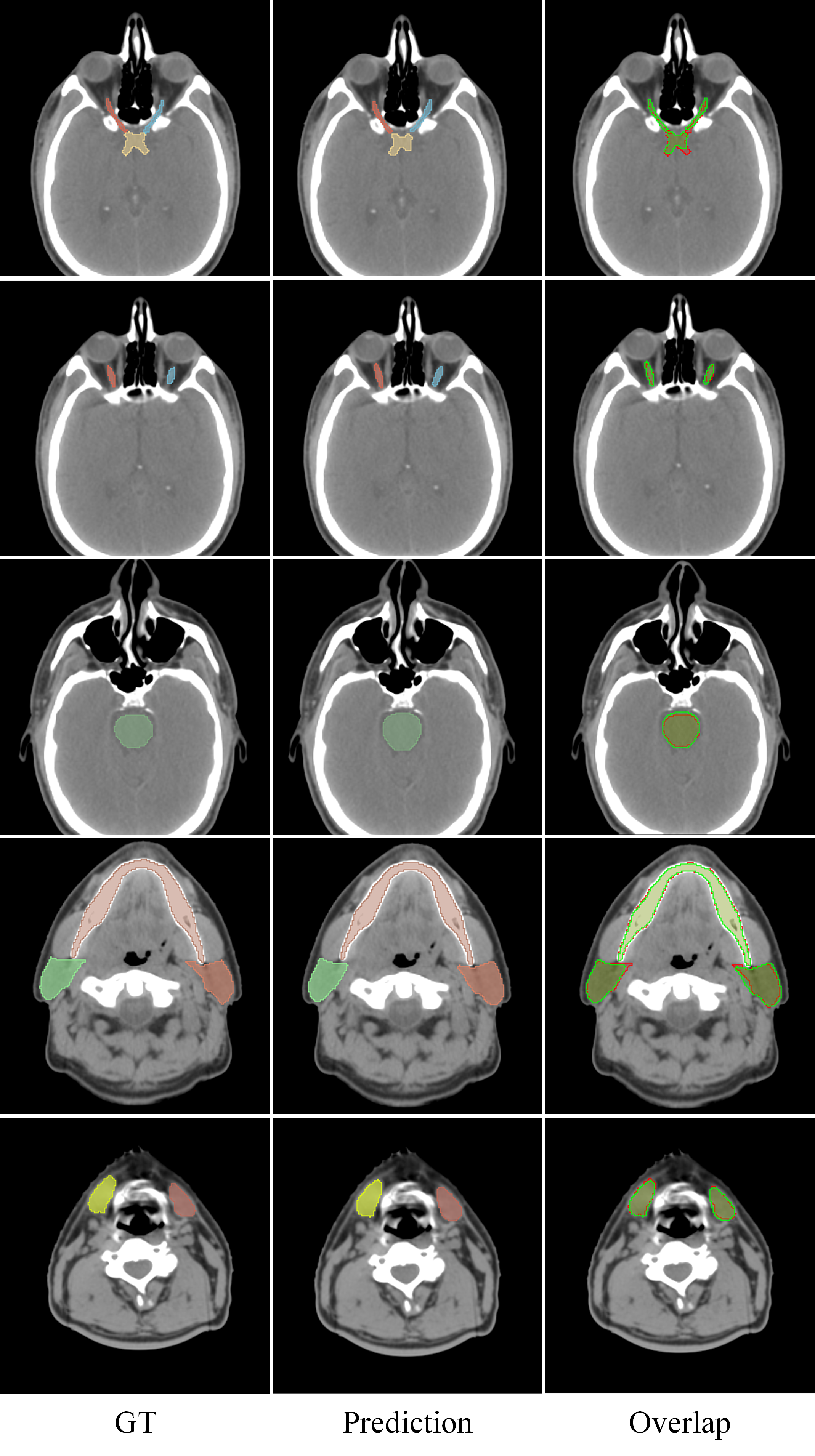}
    \caption{Visualization of results in MICCAI'15 dataset. The countour of prediction and ground truth are shown in ``Overlap''. The green lines are predictions by FocusNetv2, and the red lines are ground truth.} \label{fig:MICCAI_visual}
\end{figure}

\subsubsection{Comparison with state-of-the-arts}

We also test our FocusNetv2 on MICCAI 2015 Head and Neck dataset. All the settings of the FocusNetv2 are the same as those used in experiments on our collected dataset, except for the number of small organs is set as 3, including left and right optic nerve and optic chiasm.

We compare our method with previously published state-of-the-art results, the Dice scores and 95 HD scores are shown in Tables \ref{tab:MICCAI_dsc} and \ref{tab:MICCAI_95HD}, our method demonstrates superior overall performance, and outperforms previous state-of-the-art methods by large margins. Especially in terms of small organs, our FocusNetv2 outperforms previous state-of-the-art methods by 7.8\%. \citet{ren2018interleaved} proposed an interleaved 3D CNNs for jointly segmentation of optic nerves and chiasm. They perform patch-based segmentation in a small target volume obtained from registration. \citet{wang2018hierarchical} proposed a vertex regression-based method, which has good performance in the brain stem and mandible. However, it has relatively poorer performance in parotid glands, and they did not provide results of other organs. \citet{zhu2019anatomynet} proposed the AnatomyNet, which claims a similar idea with our backbone S-Net. But they only reduce the number of down-sampling operations, which results in limited receptive fields thus generates outliers in predicted label maps. Our S-Net introduces densely connected ASPP, which not only enlarges receptive field but also learns better multi-scale feature representations. It outputs fewer errors, as shown in 95HD in Table \ref{tab:MICCAI_95HD}.

\citet{tong2018fully} first introduced shape constraint into HaN OAR segmentation. They used an autoencoder to learn the latent shape representation from the training dataset. The main difference between our method and theirs lies in two-fold. First, we use adversarial training to learn shape representations, not only for ground truth labels, but also for estimated segmentation mask from the segmentation network. Therefore, the autoencoder can better encode subtle differences between estimated and ground-truth organ masks, thus providing more effective gradients to the segmentation network, and makes the prediction of segmentation network more consistent with ground truth shape. Second, they train the shape representation model of all organs in the original scale. However, as the extremely unbalanced ratio between background, large and small organs leads to insufficient training of small organs for the segmentation network, it will also affect the learning of shape representations by the autoencoder. Therefore, we integrate the adversarial shape loss only in our proposed SOS-Net, making the shape representation of small organs more effective. The significant performance improvement of our adversarial shape loss proves the effectiveness of our method. 

\citet{tang2019clinically} used a two-stage method, consisting of an OAR detection module and an OAR segmentation module, where the OAR detection module identifies the location and size of OAR and segmentation module is to further segment within OAR region. As the location and size of OARs are relatively consistent among patients, while using object detection module needs to carefully design anchor size and prone to have false positive, therefore, our method localizes OAR by OAR center location regression and the size of ROI is fixed based on statistics of the training set. By specially designed structure, our backbone S-Net has comparable performance with \citet{tang2019clinically}, and our final FocusNetv2 outperforms theirs by a large margin after introducing small organ segmentation branch and the proposed adversarial shape loss.

\subsubsection{Ablation studies} 

We also conduct ablation studies of the impact of each component of FocusNetv2 on MICCAI'15 dataset. The results are exhibited in Table \ref{tab:ablation_study}. We first conduct experiments on the impacts of the number of down-sampling operations in the S-Net. We train SEResUNet with 1, 2, 3 and 4 down-samplings (denoted as ``SEResUNet\_dX'') with cross-entropy loss respectively. The overall Dice score and Dice score of small organs goes up when the number of down-sampling operation decreases. The performance of SEResUNet\_d2 and SEResUNet\_d1 is similar, but SEResUNet\_d2 has slightly better performance on small organs. Utilizing the focal loss and the Dice loss (denoted as ``SEResUNet\_d2\_FL\_DL'') improves the segmentation accuracy greatly. Further combining the ASPP module into S-Net (denoted as ``SEResUNet\_d2+ASPP'') can slightly boost the performance. Introducing the small-organ localization network and the small-organ segmentation network (denoted as ``FocusNetv1''), the class imbalance problem can be solved. After adding shape constraint by adversarial autoencoder, the FocusNetv2 considerably boosts the Dice score of small organs by 5.94\%. The ``FocusNetv2 no concat'' in Table \ref{tab:ablation_study} only takes the ROI-pooled features from S-Net as the input of the small organ segmentation network (SOS-Net), where the raw CT image is not concatenated. It also adopts the AAE shape constraint. We observe that the raw CT image has a great effect on the refinement of small organ segmentation.

At last, we conduct an experiment to show the effectiveness of adversarial training of autoencoder, shown by \textit{FocusNetv2 w/ origin AE} in Table \ref{tab:ablation_study}. The autoencoder is only trained to reconstruct the inputs of ground-truth shapes from the dataset. The autoencoder's parameters are fixed to regularize the training of the segmentation network similarly to \citet{tong2018fully}, except that the regularization is only applied to the small organ segmentation branches. As the autoencoder is not trained against the segmentation network, its regularization capability is limited. The performance of \textit{FocusNetv2 w/ origin AE} is 4.11\% less than the proposed FocusNetv2, which proves the effectiveness of our adversarial autoencoder.
 
\begin{table}[htb!]

\begin{center}
\setlength{\tabcolsep}{1mm}
\begin{tabular}{lcc}
\toprule
Method                  &AVG Dice          &AVG Small Dice       \\
                       
\midrule
SEResUNet\_d4              &68.31          &38.64  \\
SEResUNet\_d3              &70.08          &41.03             \\
SEResUNet\_d2              &71.19          &45.58  \\
SEResUNet\_d1              &71.18          &44.3   \\
SEResUNet\_d2\_FL\_DL      &80.15          &66.48  \\
S-Net (SEResUNet\_d2+ASPP) &81.04          &68.71  \\
FocusNetv1                 &81.90          &71.40  \\
FocusNetv2 no concat       &82.40          &72.84  \\
FocusNetv2 w/ origin AE    &83.14          &73.23\\
FocusNetv2                 &84.51          &77.34  \\

\bottomrule
\end{tabular}
\end{center}
\caption{Ablation studies of each part of FocusNetv2. Small Dice is the average dice score of three small organs, i.e. optic chiasm, left and right optic nerve. UNet\_d4 denotes SEResUNet with 4 times of down-sampling, and so on for d3, d2, and d1, they are trained with cross-entropy loss. FL is focal loss, DL is Dice loss}\label{tab:ablation_study}
\end{table}

\subsubsection{Robustness of SOL-Net and SOS-Net} 

\begin{table}[htb!]

\begin{center}
\setlength{\tabcolsep}{5mm}
\begin{tabular}{cccc}
\toprule
Localization            &3mm           &4mm        &5mm\\
Error                   &Id. Rate      &Id. Rate   &Id. Rate\\              
\midrule
2.7 mm                  &63.3\%         &83.3\%     &100\%   \\              

\bottomrule
\end{tabular}
\end{center}
\caption{The localization results of SOL-Net on MICCAI'15 dataset. We present the localization error, and identification rate of 3mm, 4mm and 5mm. SOL-Net can identify all organs within 5mm error distance. }\label{tab:localization_error}
\end{table}

\begin{table}[htb!]

\begin{center}
\setlength{\tabcolsep}{2mm}
\begin{tabular}{ccccccc}
\toprule
SOL-Net     &0 mm       &1 mm       &3 mm       &5 mm       &7 mm       &9 mm\\             
\midrule
77.33	    &76.83	    &77.13	    &76.96	    &76.87	    &75.07	    &73.54\\
\bottomrule
\end{tabular}
\end{center}
\caption{The average Dice score of small organs with different localization error on MICCAI'15 dataset. SOL-Net means using the localization results from SOL-Net. The rests use the ground truth organ centroids, but adds random translation with distance of 0 mm, 1 mm, 3 mm, 5 mm, 7 mm and 9 mm.}\label{tab:localization_random}
\end{table}

The segmentation performance of small organs might be affected by the small-organ localization accuracy. If the organ bounding boxes deviate too far away from the ground-truth location, the SOS-Net would have difficulty on segmenting the small organs accurately. Therefore, we conducted experiments to analyze the robustness of our proposed SOL-Net and SOS-Net. First, we analyzed the localization accuracy of SOL-Net, as shown in Table \ref{tab:localization_error}. SOL-Net's average localization error (average distance between the estimated centroids and ground truth centroids) for all small organs is about 2.7mm. We further measure the small-organ localization rates within different distances. An organ is considered to be correctly localized if its estimated centroid is within a certain distance from the ground truth centroid. 63.3\% of small organs are localized within 3mm, 83.3\% can be localized within 4mm. If we extend the distance to 5mm, all small organs can be localized correctly.

Although our SOL-Net can localize small organs with small errors, we conducted another experiment to explore the effect of localization error on the following segmentation model. After the segmentation network, including S-Net, SOL-Net, SOS-Net, is trained, all parameters are fixed. Then, instead of using the localization bounding boxes from the SOL-Net, we obtain the small organs' ground truth boxes, and add random translations to simulate the localization errors. Then the randomly shifted boxes are used to guide ROI-pooling for segmentation by the SOS-Net. The results are shown in Table \ref{tab:localization_random}. For localization error within 5mm, there is a slight degradation of segmentation accuracy. Even the localization errors reach up to 9 mm, the performance degradation is still within an acceptable range. Considering that our SOL-Net's 5mm localization rate reaches 100\%,  our method is robust against the localization errors of SOL-Net.

\section{Limitations and future works}

Although our method demonstrates strong performance and can accelerate the radiotherapy planning process in clinical applications, it is still far from perfect. Our shape constraints currently only apply to small organs, as there also exists a sample imbalance problem when training the shape autoencoder. Directly training an autoencoder for all organs would favor large organs while ignoring much of the small organs, making it difficult to constrain the segmentation results of small organs. Considering that the S-Net has good performance for large organs, while there is significant variance among patients for small organs, we only apply shape constraint to small organs with blur boundaries to encourage the network to generate predictions in agreement with shape priors. Besides, our current training has multiple stages, which makes the training process complex. This is a consideration from a performance perspective. In our experiments, the network cannot achieve optimal performance when end-to-end training.

There are several main directions for future improvements. The first is to adopt novel CNN technologies, such as attention mechanisms \citep{wang2018non, fu2019dual}, to increase the capacity and capability of backbone networks. Relative position encoding \citep{dai2019transformer} can be utilized to learn relative position information between and within organs. Besides, the model's uncertainty estimation \citep{lakshminarayanan2017simple} can be provided, which is of great significance for clinical applications. It allows the doctor to pay more attention to the regions where the model is uncertain. Furthermore, our proposed adversarial shape constraints can be applied to more organs and body parts to realize the fully automatic whole-body radiotherapy planning.

\section {Conclusion}

We proposed a novel segmentation framework with adversarial shape constraint, which outperforms state-of-the-art methods on the segmentation of imbalanced OARs in HaN CT images with large margins. The framework contains two parts, a two-stage segmentation network and an adversarial autoencoder as shape regularization. The two-stage segmentation network is specifically designed for unbalanced problem of OARs in HaN CT images. By reducing the number of down-samplings and utilizing multi-scale features learned by DenseASPP, our S-Net can guarantee the accuracy of the segmentation of large organs. Trained to predict small-organ center location maps, our SOL-Net can generate accurate small-organ central locations. SOS-Net can solve the unbalanced class problem, and high-resolution feature volumes can be utilized to accurately segment small organs and thus further boost the performance. With our new adversarial autoencoder, our framework does not only learn the shape representations, but also encodes more discriminative features for organ masks. It is able to provide better supervision for training the segmentation network. Extensive experiments on a large amount of real patient data and the MICCAI 2015 dataset show the effectiveness of our proposed framework.

\section*{Acknowledgments}
This work has been supported in part by the General Research Fund through the Research Grants Council of Hong Kong under Grants CUHK14208417 and CUHK14239816, in part by the Hong Kong Innovation and Technology Support Programme (No. ITS/312/18FX), in part by NSF grants CCF-1733843.

\bibliographystyle{model2-names.bst}\biboptions{authoryear}
\bibliography{refs}

\section*{Appendix}

\subsection{Network details}

In this section, we provide more details about our network structure, including the segmentation network and the autoencoder.

The structure of autoencoder is shown in Table \ref{tab:ae_structure}, which is a convolutional autoencoder. Batch normalization is also used before each ReLU layer except for the FC layer. We use trilinear upsampling layers follows convolution layers instead of transpose convolution layers in the decoder. The output of the first FC layer, which is a 512 dimentional feature vector, is the latent code of the autoencoder.

The structure of segmentation network is shown in Table \ref{tab:segnet_structure}, including S-Net, SOL-Net, and SOS-Net. For SOS-Net, the first maxpooling uses kernel size of (1,2,2) as the spacing of CT images is anisotropy, and is larger in z-axis. In the decoder, SOS-Net uses trilinear upsampling layers follow SEResBlocks to recover the spatial resolution.

\begin{table}[h]

\begin{center}
\setlength{\tabcolsep}{2mm}
\begin{tabular}{cccccc}
\toprule
        &Kernel     &Stride/Scale       &Channels       &None-linear       \\      
\midrule
Conv3D  &5	        &2	                &64	            &ReLU\\
Conv3D  &5          &2                  &128            &ReLU\\
Conv3D  &5          &1                  &128            &ReLU\\
Conv3D  &5          &2                  &256            &ReLU\\
FC      &-          &-                  &512            &None\\
FC      &-          &-                  &256            &ReLU\\
Upsample&-          &2                  &-              &-  \\
Conv3D  &5          &1                  &128            &ReLU\\
Conv3D  &5          &1                  &128            &ReLU\\
Upsample&-          &2                  &-              &-  \\
Conv3D  &5          &1                  &64             &ReLU\\
Upsample&-          &2                  &-              &-  \\
Conv3D  &5          &1                  &1              &ReLU\\
\bottomrule
\end{tabular}
\end{center}
\caption{Structure of the adversarial autoencoder model}\label{tab:ae_structure}
\end{table}

\begin{table}[h]

\begin{center}
\setlength{\tabcolsep}{1mm}
\begin{tabular}{clc}
\toprule
                    &layer name                                 &Channels        \\      
\midrule
S-Net               &Conv3D stride=1, kernel size=3              &32	        \\
                    &SEResBlock $\times 1$                      &32         \\
                    &maxpooling kernel size=(1,2,2)             &-          \\
                    &SEResBlock $\times 2$                      &48         \\
                    &maxpooling kernel size=(2,2,2)             &-          \\
                    &SEResBlock $\times 2$                      &64         \\
                    &SEResBlock $\times 2$, dilation rate=2     &64         \\
                    &DenseASPP  dilation rate=3,6,12,18         &320        \\
                    &Upsample   scale=(2,2,2)                   &-          \\
                    &SEResBlock $\times 1$                      &48         \\
                    &Upsample   scale=(1,2,2)                   &-          \\
                    &SEResBlock $\times 1$                      &32         \\
                    &Conv3D stride=1, kernel size=1             &1          \\
\midrule
SOL-Net             &SEResBlock $\times 1$                      &32         \\
                    &Conv3D stride=1, kernel size=1             &Num small organs    \\
\midrule
SOS-Net             &SEResBlock $\times 2$                      &32         \\
                    &Conv3D stride=1, kernel size=1             &1          \\
                    
\bottomrule
\end{tabular}
\end{center}
\caption{Structure of the segmentation network}\label{tab:segnet_structure}
\end{table}

\subsection{More visualization results}
\begin{figure}[h]
\centering
    \includegraphics[height=13cm]{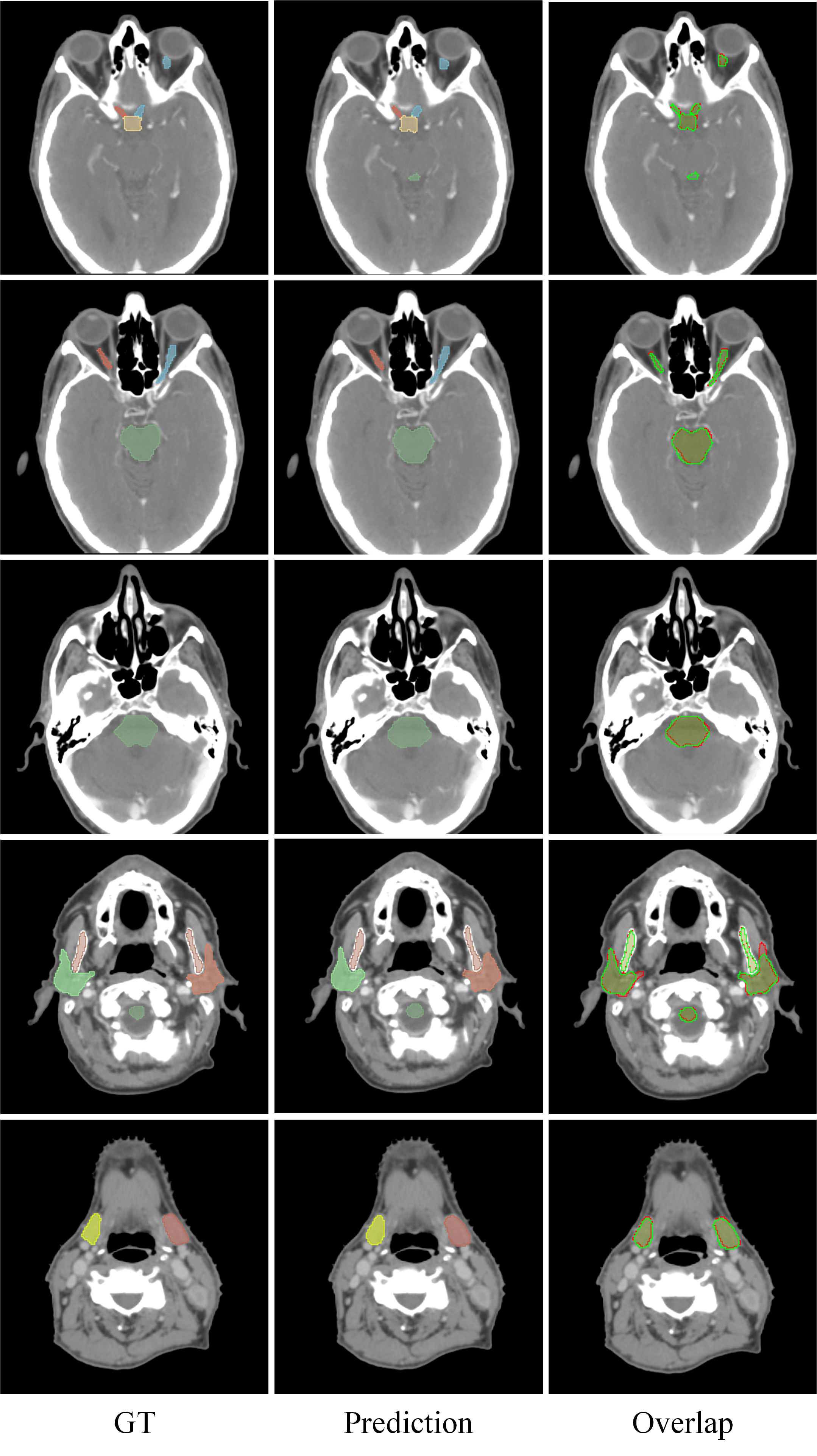}
    \caption{Supplementary figure. More visualization of results in MICCAI2015 dataset. The countour of prediction and ground truth are shown in ''Overlap'', the green line is FocusNetv2 prediction, the red line is ground truth.} \label{fig:more_MICCAI_visual}
\end{figure}

\end{document}